\documentclass[10pt,twocolumn,journal]{IEEEtran}

\usepackage[T1]{fontenc}% optional T1 font encoding

\usepackage{amsmath}
\usepackage{setspace}
\usepackage{amssymb}
\usepackage{stfloats}
\usepackage{cite}
\usepackage{ragged2e}
\usepackage[ruled,vlined,linesnumbered]{algorithm2e}
\usepackage{amsfonts}
\usepackage{mathrsfs}
\usepackage{amsmath,amsthm}
\usepackage{array,booktabs}
\usepackage{subfigure}
\usepackage{multirow}
\usepackage{cuted}
\usepackage{multicol}
\usepackage{graphicx}
\usepackage{subfigure}
\usepackage{graphicx,xcolor,bm}
\usepackage{hyperref}
\usepackage{threeparttable}
\usepackage{dcolumn}
\usepackage{setspace}
\usepackage{makecell}
\usepackage{lipsum}
\usepackage{cite,bm}
\graphicspath{{figures/}}
\begin{document}
%
%\title{Overbook in Advance, Trade in Future (OATF): A Hybrid Resource Overbooking Mechanism in Device-Edge-Cloud Networks}

\title{Overbook in Advance, Trade in Future: \\ Computing Resource Provisioning in Hybrid Device-Edge-Cloud Networks}
%Overbook in Advance, Trade in Future
\author{Minghui Liwang, \IEEEmembership{Member, IEEE}, and Xianbin Wang, \IEEEmembership{Fellow, IEEE}
\thanks{Minghui Liwang is with Xiamen University, Xiamen, China. Xianbin Wang is with Western University, Ontario, Canada. 

Corresponding author: Minghui Liwang, Xianbin Wang. 

E-mail: minghuilw@xmu.edu.cn, xianbin.wang@uwo.ca.
}}
\maketitle
\newcommand\blfootnote[1]{%
\begingroup
\renewcommand\thefootnote{}\footnote{#1}%
\addtocounter{footnote}{-1}%
\endgroup
}
%\IEEEtitleabstractindextext{
\begin{abstract}
%\justifying
The big data processing in distributed Internet of Things (IoT) systems calls for innovative computing architectures and resource provisioning techniques to support real-time and cost-effective computing services. This article introduces a novel overbooking-promoted forward trading mechanism named \textit{Overbook in Advance}, \textit{Trade in Future} (OATF), where computing resources can be traded across three parties, i.e. end-users, an edge server and a remote cloud server, under a hybrid device-edge-cloud network with uncertainties (e.g., ``no shows''). More importantly, OATF encourages a feasible overbooking rate that allows the edge server to overbook resources to multiple end-users (e.g., exceed the resource supply), while purchasing backup resources from the cloud server, by determining rights and obligations associated with forward contracts in advance via analyzing historical statistics (e.g., network, resource dynamics). Such a mechanism can greatly improve time efficiency and resource utilization thanks to overbooking and pre-signed forward contracts. Critical issues such as overbooking rate design and risk management are carefully investigated in this article, while an interesting case study is proposed with mathematical analysis. Comprehensive simulations demonstrate that OATF achieves mutually beneficial utilities for different parties (cloud, edge, and end-users), as well as substantial resource usage and commendable time efficiency, in comparison with conventional trading methods. \blfootnote{}
\end{abstract}

%}
%}

% \fi
%\IEEEpeerreviewmaketitle

\section{Introduction}
\IEEEPARstart{R}{ecent} years have witnessed the rapid evolution of communication technologies and fast proliferation of Internet of Things (IoT) devices with growing capabilities of data gathering, analyzing and knowledge utilization assisted by wireless networks~\cite{1,2,3,4}, which also enable a wide range of advanced applications, e.g., smart factory, intelligent transportation, augmented reality (AR)/virtual reality (VR) games, etc. However, constrained computing resources and capabilities of IoT devices pose great challenges in handling ever-growing computation-intensive and time-sensitive application data~\cite{5}. Besides, limited power and battery supply~\cite{6} of smart devices may further prevent smooth on-board application processing. Such challenges urgently call for cost-effective, reliable and efficient computing resource provisioning techniques over connected IoT systems to secure necessary resources for computation-intensive applications.
\vfill

\subsection{Motivations}

%tab1
\begin{table*}[t!]
%\setstretch{0.9}
%{\footnotesize
%\small
%\setstretch{0.9}
%\renewcommand{\arraystretch}{1.3}
\caption{Comparison among different resource trading modes}
\label{table_example}
\centering
\setlength{\tabcolsep}{7mm}{
\begin{tabular}{|c|c|c|c|}
\hline
 \textbf{Resource trading mode} & \textbf{Spot trading} & \textbf{Conventional booking} & \textbf{Overbooking}\\ \hline
Data basis & Current statistics & Historical statistics  & Historical statistics \\ \hline
Decision-making overhead ($d$) &\multicolumn{3}{|c|}{~$d\left(\text{Overbooking}\right)$~$\leq $~$d\left(\text{Conventional booking}\right)$~$<$~$d\left(\text{Spot trading}\right)$}\\ \hline
Time efficiency ($t$) &\multicolumn{3}{|c|}{$t\left(\text{Overbooking}\right)$~$>$~$t\left(\text{Conventional booking}\right)$~$>$~$t\left(\text{Spot trading}\right)$ }\\ \hline
Resource utilization rate ($r$) &\multicolumn{3}{|c|}{$r\left(\text{Conventional booking}\right)$~$\leq$~$r\left(\text{Overbooking}\right)$~$\leq$~$r\left(\text{Spot trading}\right)$}\\ \hline
\end{tabular}}
%}
\end{table*}

\noindent
This article studies a novel overbooking-promoted resource provisioning paradigm under hybrid device-edge-cloud network architecture~\cite{7,8,9}, in supporting effective and reliable computing services. The following key questions have been identified, which represent our major motivations. 

%\noindent
%$\bullet$~
%\textit{(i). Why resource provisioning under hybrid device-edge-cloud network architecture is needed in supporting IoT applications?}
%Although cloud computing offers valuable resources to IoT devices, which, however, inevitably incurs excessive transmission latency and heavy burden on network infrastructure, and may be inappropriate in handling massive amount of distributed data in many IoT applications~\cite{7}. 

%Edge computing can provide cloud-like computing/storage capacities at the network edge, while thus facilitating more effective resource provisioning and responsive computing services~\cite{7}. Nevertheless, edge-driven networks alone may face significant shortage of computing/storage resources when meeting ever-growing resource demands from dramatic growing number of smart devices. 

%that allows smart devices to rent and use computing resources in supporting complicated mobile applications
%\noindent

%\noindent
%$\bullet$~
%\textit{(ii). Why resource booking is essential?} Since incentive presents a critical factor in facilitating consensual and robust resource sharing mainly due to the selfishness of participants, which generally require a beneficial resource trading arrangement. For instance, each end-user can offload a certain amount of application data to edge server for execution by getting access to a nearby access point (AP, e.g, base stations, etc.), via paying for the obtained resources and computing services. Besides, the edge server will be charged for purchasing resources from the remote cloud server. 

\textit{(i). Why computing resources should be booked in advance?} Mutually beneficial resource trading mechanisms are the foundation of distributed computing resource sharing due to the selfishness of every participant. Consequently, incentives for resource sharing plays a critical role in facilitating consensual and reliable resource provisioning among multiple parties. For instance, an end-user can offload a certain amount of application data to edge server for execution by getting access to a nearby access point (AP, e.g, base stations, etc.), via paying for the obtained resources and computing services. At the same time, the edge server could be charged for purchasing resources from a remote cloud server. 

In securing necessary distributive resources, conventional onsite spot trading presents a widely adopted paradigm that enables resource selling/buying among resource owners and requesters according to the current system conditions (e.g., resource supply/demand and wireless channel qualities between end-users and APs at present), which, however, can cause significant performance degradation, e.g., overhead can be incurred for discussing/negotiating the final trading agreement~\cite{3,10,4} which thus can lead to unsatisfying time/resource efficiency. Take online auction as an example, computing resources that have been put aside for unsuccessful trading during the decision-making procedure may cause resource underutilization. Moreover, spot trading participants are generally risking failures to access the required resources, e.g., only a finite number of winners can finally obtain limited resources during an onsite auction, while losers receive nothing even though they have spent both time and energy on bidding/waiting/negotiating during the auction procedure. This case can further result in unsatisfying trading experience~\cite{4,11}.
%For instance, a smart device who is supposed to process its application locally at time $t$, can only start processing at $t+\Delta t$ when he fails in a trading; namely, $\Delta t$ has been somehow wasted during onsite decision-making. 
Since the big data generated by massive IoT devices often expects real-time processing, the above-mentioned shortcomings prompt the authors to investigate efficient resource trading mechanisms. To this end, \textit{resource booking} is considered in this article which facilitates a \textit{forward trading} manner (namely, presale), where a resource owner and a requester can reach an agreement for future practical trading in advance~\cite{3,10}, via signing a forward contract associated with contract terms, such as reasonable resource price, the amount of trading resources, and default clause if either party breaks the contract, etc. The benefit of the pre-signed trading contracts is that participants will no longer have to spend extra time/energy on onsite decision-making, which can thus improve time efficiency. An example of timeline comparison is depicted by Fig. 1, where in forward trading, the actual service can be directly delivered without any negotiation thanks to pre-signed contracts. 

%f1
\begin{figure}[h!t]
\centerline{\includegraphics[width=1\linewidth]{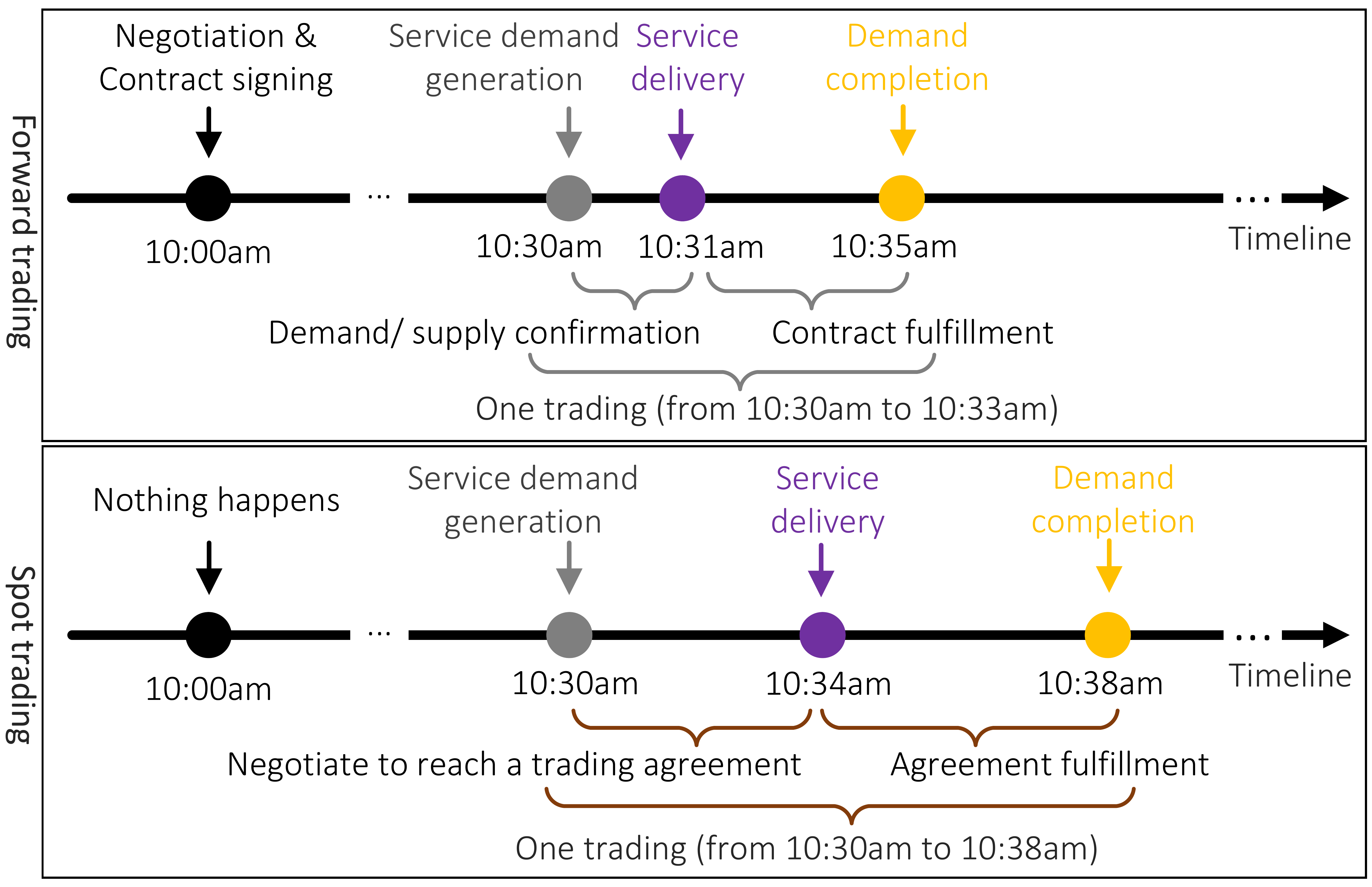}}
\caption{Timeline comparison associated with spot trading and forward trading.}
\label{fig1}
\end{figure}

\textit{(ii). Why resource overbooking is critical in dynamic networks?} Conventional resource booking mechanism that allows the equal amount of reserved (booked) and available resources for requesters is generally ineffective in handling networks with dynamic resource demands. This is provoked by factors such as uncertain mobility and willingness of requester, varying wireless communication conditions, etc. For example, "\textit{no shows}" of requesters are common in real-life networks, where smart devices that lose connections with the edge server, or have run out of power, will not participate in a trading and thus prevent the originally confirmed utilization of booked resources as stipulated in pre-signed contracts. This case can further incur the underutilization of dynamic resources.
To achieve better resource usage, \textit{overbooking}~\cite{12,13} has been introduced as a common practice in many fields (e.g., airlines and hotels, spectrum reservation, etc.), encouraging the promissory reserved resources in excess of actual resource supply owing to dynamic demands. For example, airlines routinely overbook flight tickets by ensuring the maximum number of travelers since some of them may be absent from the planned trip; otherwise, each flight usually takes off with roughly 15\% seats empty, which further incurs nonnegligible economic losses\cite{11, 12}.
Similarly, encouraging proper overbooking rate in computing resource trading market can greatly support substantial resource utilization and profit advantages, via analyzing historical statistics associated with the uncertain resource demand and supply (e.g., uncertain user’s participation, varying channel quality).

\textit{(iii). Why resource overbooking in hybrid device-edge-cloud networks is challenging?}
Integrating cloud and edge into a hybrid computing system represents a viable solution~\cite{8,9} to overcome the possible resource shortage of the edge server, where the remote cloud server 
plays the role of a powerful backup resource supply center, in supporting more end-users and applications while attracting better profits. Although device-edge-cloud network architecture efficiently unifies distributed heterogeneous resources for service provisioning, additional challenges would be incurred. For example, since resources should be overbooked across three parties: end-users, edge server, and cloud server, the amount of resources that the edge decides to purchase from the cloud server relies heavily on the dynamic resource demand of end-users, where an inappropriate overbooking rate can result in performance degradation for computing service delivery. Besides, the cloud server generally has to serve other requesters, where the uncertain resource supply can leave impacts on the cloud server's willingness to sell resources to the edge server. 
Thus, dynamic resource supply/demand, and the individual rationality associated with different parties in hybrid device-edge-cloud networks greatly call for designing feasible resource overbooking rate. For example, the overbooking rate should be beneficial to different parties, which considers both the overbooking procedure among end-users and edge, as well as that between edge and cloud. The above discussions represent the most significant motivations. Specifically, Table 1 shows the conclusive differences among different resource trading methods on critical evaluation indicators. 

%f2
\begin{figure*}[t!]
\centerline{\includegraphics[width=1\linewidth]{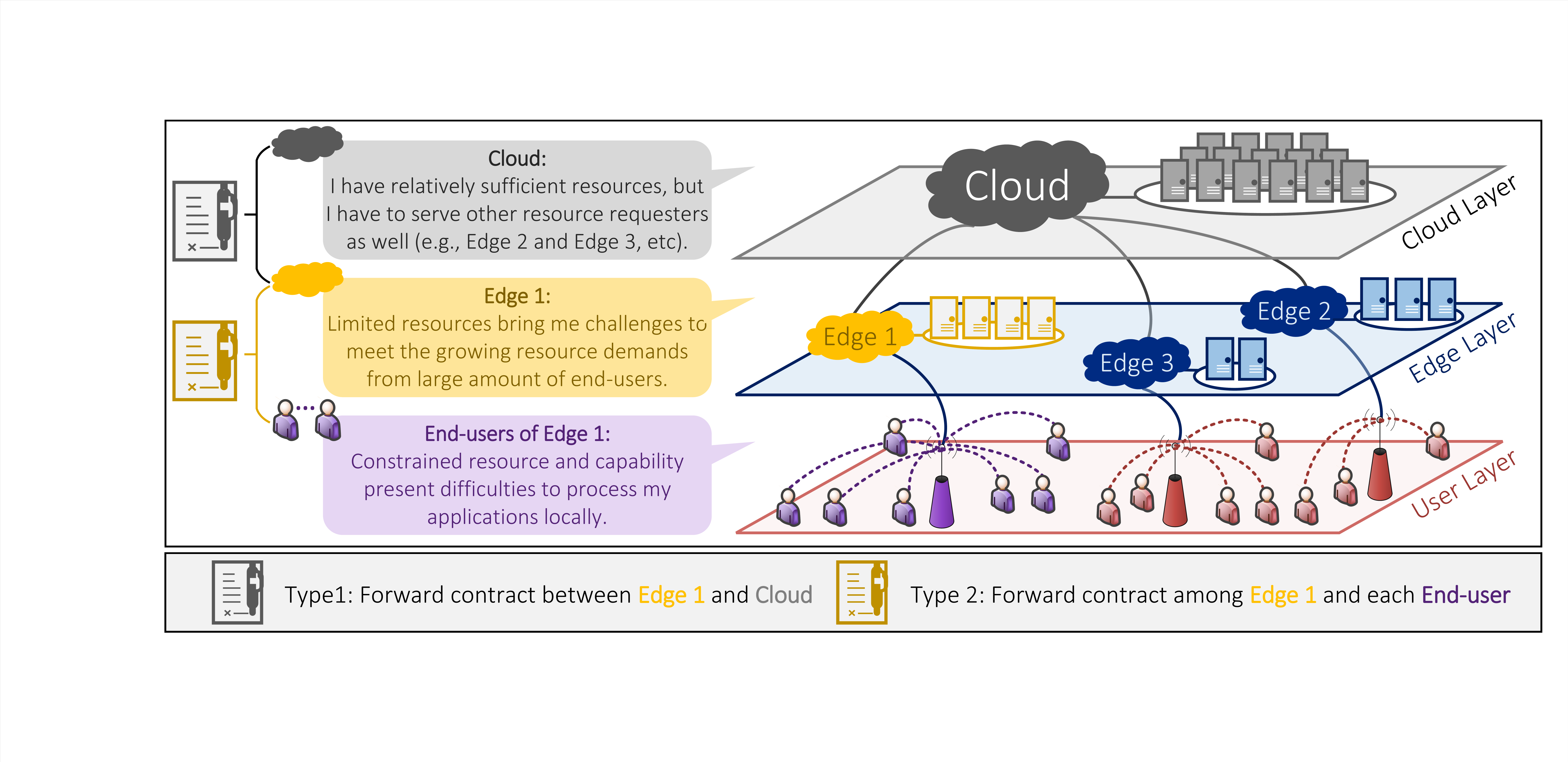}}
\caption{Framework of the proposed OATF under hybrid device-edge-cloud network architecture, where this article mainly relies on investigating the resource trading among Cloud server, Edge 1, and end-users associated with Edge 1.}
\label{fig2}
\end{figure*}

%\vfill

\subsection{Contributions}
\noindent
Most existing works mainly focus on spot trading \cite{8,15} or conventional booking\cite{3}, which are facing challenges in handling dynamic and unpredictable nature of resource trading market. This article proposes a novel overbooking-promoted trading mechanism for computing resources named ``\textit{Overbook in Advance, Trade in Future}'' (OATF), under device-edge-cloud network architecture, which contains three layers: user layer, an edge layer, and cloud layer (namely, three different parties). This article aims to investigate comprehensive insights on how the proposed OATF mechanism can facilitate efficient resource provisioning associated with end-users, an edge server, and a cloud server. Major contributions are summarized below:

%\textit{i)} users layer, where each end-user may have multiple applications need to be either processed locally or offloaded to computing servers; \textit{ii)} edge layer, where the edge server is facing challenges to meet massive resource demands from end-users due to constrained resource supply; and \textit{iii)} cloud layer, where the cloud server acts as an applicable backup resource supply center. 

\noindent
$\bullet$ OATF, a novel overbooking-enabled forward trading mechanism for computing service, is proposed under device-edge-cloud network architecture. Various uncertainties are considered to capture the random and unpredictable nature of resource trading market. Specifically, the edge server can overbook resources to multiple end-users while purchasing backup resources from the cloud server, by signing forward contracts in advance, via analyzing historical trading statistics. The overall framework and relevant key issues, e.g., contract term determination and risk evaluation, are analyzed in detail.

\noindent
$\bullet$ A case study is investigated to describe how the proposed OATF mechanism can be implemented in practice. For which, a multi-objective optimization (MOO) problem is formulated, while a two-way multilateral negotiation scheme is designed that facilitates the trading among different parties.

\noindent
$\bullet$ Comprehensive experimental results demonstrate that the proposed OATF mechanism achieves commendable benefits for participants from three different parties, while outperforming benchmark methods on critical evaluation factors, e.g., application completion time, undesired trading failure, time-efficiency, and resource utilization.

%This article is structured as follows: the framework and key issues associated with the proposed OATF mechanism under device-edge-cloud networks are firstly introduced. Then, an interesting case study is investigated which mainly describes how OATF can be implemented, where the experimental results demonstrate that OATF achieves commendable performance as compared with benchmark methods. Finally, we conclude this article before analyzing interesting future research directions. 

%\vfill

%2
\section{Overbook in Advance, Trade in Future}

\subsection{Overview}
\noindent
The hybrid device-edge-cloud network architecture contains three key layers: user layer, edge layer, and cloud layer, as illustrated in Fig. 2.

\noindent
\textbf{User Layer} mainly includes smart devices (e.g., smartphone, smart vehicle, drone, etc.) with intelligent applications, which, however, are facing difficulties to process application data locally, owing to limited on-board computing/storage resources and capabilities. Fortunately, this framework allows end-users to purchase resources and computing services from edge server (or cloud server) by offloading a certain amount of application data via getting access to nearby APs. Notably, APs are connected to edge servers through fiber-optic links \cite{14}.

\noindent
\textbf{Edge Layer} is generally composed of several edge servers close to end-users, which can offer computing services under cost-effective and responsive manner. However, the limited computing/storage resource supply of a single edge server brings challenges to meet the ever-growing resource demands, mainly incurred by the big data generated on countless IoT devices and the wide range of innovative mobile applications. Thus, an edge server may have to purchase more resources from the remote cloud server especially during peak hours. Specifically, edge servers are connected to the cloud server via fiber-optic links\cite{14}.

%f3
\begin{figure*}[t!]
\centering
\centerline{\includegraphics[width=0.99\linewidth]{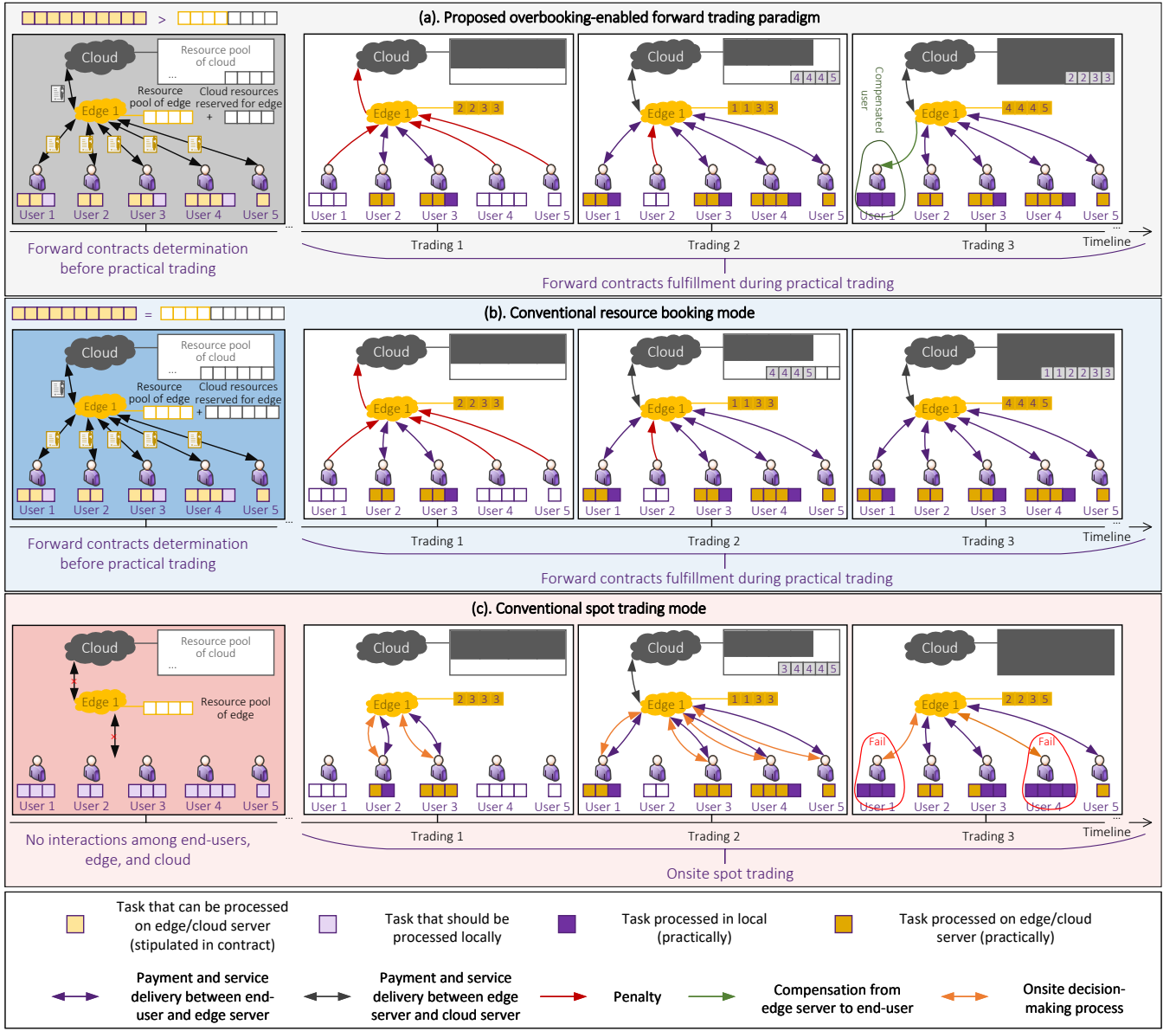}}
\caption{Timelines and trading examples upon comparing different resource trading modes.}
\end{figure*}

\noindent
\textbf{Cloud layer} considers a remote cloud server distant from end-users as a relatively powerful data/computing center, which provides highly precise computing service for mobile applications. However, due to the potential excessive transmission delay and burdens on wireless links as well as cloud server, direct communications among end-users and cloud server are generally not recommended. Instead, the cloud server is seen as an effective backup resource supply center that lends resources to edge servers, and thus helping with attracting more end-users and revenue.

Resource trading considered in this article mainly investigates the interactions among the cloud server, one edge server (e.g., Edge 1 in Fig. 2), and the relevant end-users (e.g., end-users of Edge 1 in Fig. 2). Note that a resource market under mobile wireless communication networks is always dynamic, inherent uncertainties should thus be carefully concerned from two key perspectives: resource demand/supply, and network condition. The uncertainty associated with resource demand mainly refers to the fluctuating number of applications, as well as "show/no show" cases of end-users. Considering a resource trading at time $t$, end-users may carry different number of heterogeneous applications, which directly impacts the amount of required resources. Besides, end-users may not always participate in a trading, e.g., a smart device outside of computing server's communication coverage or has run out of power will be absent from time $t$ and thus not using the booked resources (namely, ``no show''). Uncertain resource supply generally depends on the cloud server, since it may have to offer services to many other customers. For instance, in Fig. 2, the amount of cloud resources provisioned to Edge 2 and Edge 3 can directly affect the available resources for Edge 1. Then, the uncertain network condition is mainly reflected by varying wireless channel qualities among edge server (namely, APs) and end-users, incurred by factors such as users' mobility and transmission power, etc. Apparently, a poor channel quality poses significant impact on application execution performance, e.g., a large data transmission delay.

By analyzing historical statistics associated with the above-mentioned uncertainties, the proposed OATF mechanism encourages two forward trading contract types (see Fig. 2): \textit{Type 1} indicates the forward contract among end-users and the edge server; and \textit{Type 2} represents the forward contract between edge server and the cloud server. Particularly, every practical resource trading is performed among participants depending on pre-determined forward contracts without further onsite negotiation. Specifically, aiming to achieve substantial utilization and profit advantages under dynamic resource supply/demand, a certain overbooking rate is encouraged, which allows the amount of booked resources stipulated in forward contracts to exceed the available resource supply. For example, the total promissory reserved resources for end-users $r^{User}$ can be larger than the available resource supply $r^{Edge}+r^{Backup}$, where $r^{Edge}$ and $r^{Backup}$ denote the edge server's local resources, and the available backup resources borrowed from cloud server, respectively.

%\subsection{Key Issues}
\subsection{Significant Issues}
Timeline associated with the proposed OATF can generally be divided into two phases: before practical trading, and during practical trading, where significant issues of the former phase are analyzed below.

\noindent
$\bullet$\textbf{Contract term design (rights and obligations)}: By signing a forward contract with the edge server, each end-user can enjoy the following rights during practical trading: \textit{(i)} a certain amount of reserved resources; \textit{(ii)} reasonable price for trading resources; and \textit{(iii)} a compensation from the edge server if the end-user fails to acquire the promissory resources owing to insufficient resource supply. Besides, each contractual end-user also has to follow the obligation by paying a certain penalty to edge server when it is absent from a trading (namely, breaks the contract). 
Apparently, the above-mentioned \textit{(i)}-\textit{(iii)} are obligations of the edge server, while the penalty paid from end-users stands for its right in contract type 1. It is noteworthy that since each contract involves one specific user, the user selection problem can be figured out accordingly. Besides, edge server can also purchase computing services from the cloud server to meet the growing resource demands of end-users, by enjoying a certain amount of reserved cloud resources, proper trading price, and a compensation when the cloud server breaks the contract since it also has to serve other customers (namely, edge's rights, cloud's obligations associated with contract type 2). Similarly, edge server has to pay a penalty for not buying the confirmed cloud resources (namely, edge's obligation, cloud's right associated with contract type 2). For example, the edge's local resources may be sufficient to cover resource demands when few end-users participate in a trading. More importantly, our proposed OATF greatly supports fairness, since the pre-determined prices will not be impacted by the uncertainties in resource trading market. Apparently, inappropriate rights and obligations associated with different forward contracts can definitely bring performance degradations, e.g., large resource price/penalty may lead to negative utilities to end-users. Thus, designing feasible contract terms represents a significant problem. 

%f4
\begin{figure*}[t!]
\centering
\centerline{\includegraphics[width=1\linewidth]{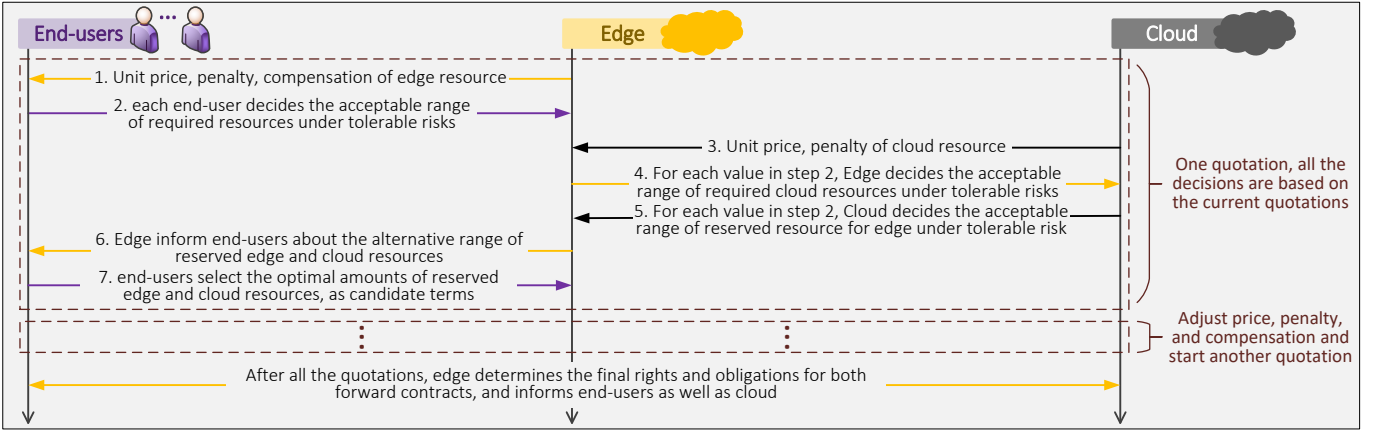}}
\caption{Procedure of the proposed negotiation scheme for contract design.}
\end{figure*}

\noindent
$\bullet$ \textbf{Overbooking rate design}: Overbooking rate refers to the proportion of promissory reserved resources that exceeds the available resource supply, e.g., $\frac{r^{User}-\left(r^{Edge}+r^{Backup}\right)}{r^{Edge}+r^{Backup}}$. Infeasible overbooking rate may incur two major problems: \textit{(i)} a large value of overbooking rate, namely, overmuch promissory reserved resources, can prevent some users from enjoying computing service owing to limited resource supply, and thus results in poor trading experience; and \textit{(ii)} a small value of overbooking rate, namely, deficient resources available for booking, can lead to underutilization of dynamic resources and further bring economic losses to computing servers. Besides, additional challenges can be incurred since overbooking is considered across three layers, where the dynamic resource demands from end-users can definitely impact the amount of trading resources of both two contracts. Consequently, overbooking rate should be well designed by comprehensively analyzing historical statistics of various uncertainties. 

\noindent
$\bullet$ \textbf{Risk management}: Uncertainties can generally bring risks, mainly in forms of \textit{(i)} participants' utilities, and \textit{(ii)} resource usage. The former indicates that participants are at risk of obtaining undesired utilities during each trading. For example, a contractual end-user who is suffering from a poor wireless channel quality and a high pre-determined resource price may receive negative utility during a practical trading, due to excessive data transmission delay and large payment. The edge server may get unsatisfying utility for paying high penalty to cloud server when lots of end-users are absent from a trading. Besides, since the cloud server generally offers services to multiple customers (e.g., Edges 2-3 in Fig. 2), a large amount of reserved resources for the concerned edge server (Edge 1) can directly reduce the resource supply and thus the relevant revenue. The later risk is mainly caused by overbooking, where a contractual end-user may still fail to access the required resources. Although he gets compensation, this case can definitely lead to poor trading experience. Additionally, the edge server is risking inadequate resource usage rate due to possible "no shows". Therefore, risks should be properly managed and controlled.

During each practical trading (the later phase), the following key issues should be carefully considered.

\noindent
$\bullet$ \textbf{Compensated user selection}: Limited resource supply can incur the case where end-users who have signed contracts but finally fail to acquire resources, and should process applications locally. Since different end-users may have heterogeneous applications (e.g., data size, etc.) and requirements (e.g., tolerant completion time, etc.), a proper selection strategy should be concerned for choosing appropriate compensated end-users (if any), during each practical trading. Common methods can refer to first-come-first-serve (FCFS), random selection (e.g., users are randomly be compensated), greedy-based selection (e.g., users with the worst channel qualities will be compensated), etc. 

\noindent
$\bullet$ \textbf{Application transfer decision}: 
Note that different computing servers can provide heterogeneous resources and services. Thus, which application(s) of which end-user(s) could be transferred to the cloud server, or stay on the edge server, presents another noteworthy problem, since cloud can offer rather powerful computing capability and may directly impact the performance of applications. Factors such as the tolerant delay of mobile application, preference of end-user, distance and channel qualities between users and APs have to be taken into consideration.

Fig. 3 shows the timeline and trading examples related to the proposed OATF mechanism, in comparison with conventional booking method (e.g., equal-booking-related trading~\cite{4}) and spot trading. Apparently, forward contracts are pre-signed among participants in Fig. 3(a) and Fig. 3(b), where contractual users will no longer spend extra time and energy on decision-making during each practical trading. Specifically, OATF in Fig. 3(a) encourages a certain overbooking rate calculated by $(10-8)/8=25\%$ in case of possible "no shows".  Fig. 3(b) depicts the conventional booking method where the overall resources booked to end-users can not exceed resource supply. For example, in Trading 2, Fig. 3(b), where user 2 is absent, the resource utilization is calculated by $80\%$; while that of our proposed OATF achieves $100\%$ (see Fig. 3(a)) which thus can better deal with dynamic resource demands. As a comparison, Fig. 3(c) illustrates the onsite spot trading mode, where the actual data offloading and service delivery procedure can only start after an onsite trading agreement has been reached.  This case can definitely lead to extra latency and energy costs incurred by onsite decision-making. Besides, undesired trading failures might be incurred. For example, in Trading 3 (Fig. 3(c)), although users 1 and 4 have spent a certain amount of time negotiating the trading agreement with the edge server, they finally fail to obtain the required service due to resource shortage. 
 
\section{Case Study}

This section investigates a case study associated with OATF upon considering three key parties: multiple end-users, an edge server with $r^{Edge}$ resources, and a cloud server with $r^{Cloud}$ resources. Namely, edge server and cloud server can theoretically process a maximum of $r^{Edge}+r^{Cloud}$ applications in parallel during a trading, for analytical simplicity.

\subsection{Basic Modeling}

Considering multiple independent identically distributed (i.i.d) end-users $\mathbb{U}=\left\{u_x\left|x\in\right.\left\{1,2,...,|\mathbb{U}|\right\}\right\}$ which are supposed to have same computing capability (e.g., similar smartphone types with the same processors), number of applications $n$, etc., for analytical simplicity. Consequently, terms of contract $\mathbb{M}^{Edge}$ offered by the edge server can thus be the same among different users, which is also general in real-life networks~\cite{4}. Specifically, each user $u_x$ may encounter two uncertain factors: $\alpha_x$ and $\gamma_x$, where random variable $\alpha_x$ defines the attendance 
%($\alpha_x=1$ at probability $a$)
and absence 
% ($\alpha_x=0$ at probability $1-a$) 
of $u_x$ during a trading, that obeys a Bernoulli distribution represented by $\alpha_x\sim\text{B}\left\{\left(1,0\right),\left(a,1-a\right)\right\}$. Besides, random variable $\gamma_x$ describes the changing channel quality of the link between $u_x$ and the nearby AP, which follows a uniform distribution denoted by $\gamma_x\sim \text{U}\left(y_1,y_2\right)$, where a small value of $\gamma_x$ can lead to excessive transmission latency of application data. Specifically, contract $\mathbb{M}^{Edge}$ offered by edge server is denoted as a tetrad,
%$\mathbb{M}^{Edge}=\left\{r^{user},p^{UtoE},q^{UtoE},c^{EtoU}\right\}$, 
where the utility $\mathcal{U}^{u_x}$ of an end-user $u_x$ who has signed contract $\mathbb{M}^{Edge}$ considers the following factors: \textit{(i)} the saved time and energy as benefited from the amount of reserved resources $r^{user}$; \textit{(ii)} payment to edge server $p^{UtoE}$ for required resources and service; \textit{(iii)} penalty to edge server $q^{UtoE}$ for possible absence; and \textit{(iv)} possible compensation $c^{EtoU}$ obtained from edge server.
%\mathbb{M}^{Cloud}=\left\{r^{Backup},p^{EtoC},q^{EtoC}\right\}
Let $\mathbb{M}^{Cloud}$ indicate the forward contract between cloud server and edge server. Relying on both $\mathbb{M}^{Edge}$ and $\mathbb{M}^{Cloud}$, the utility of edge server $\mathcal{U}^{Edge}$ concerns four parts: \textit{(i)}. revenue obtained from end-users; \textit{(ii)}. compensation that edge should pay to the end-users who fail to acquire resources; \textit{(iii)}. payment $p^{EtoC}$ for the predetermined amount of cloud resources $r^{Backup}$; and \textit{(iv)}. penalty $q^{EtoC}$ if not purchasing cloud resources. 

This case study considers an interesting assumption that the cloud server will offer services to the studied edge server as the highest priority (namely, the cloud server will not break $\mathbb{M}^{Cloud}$). Accordingly, the uncertain resource demand from other customers obeys a discrete uniform distribution denoted by $\beta\sim\text{U}\left(0,1,...,r^{Cloud}\right)$. Since the pre-determined contract $\mathbb{M}^{Cloud}$ has set aside $r^{Backup}$ resources for the studied edge server, some of these requesters may have to wait for resource release during a trading when cloud server is fully occupied.
%(e.g., $r^{Cloud}-r^{Backup}>\beta$). 
Consequently, utility $\mathcal{U}^{Cloud}$ of the cloud server is defined via considering four key parts: \textit{(i)} revenue obtained from other resource requesters; \textit{(ii)} partial refund for requesters who have to wait for available resources; \textit{(iii)} income $p^{EtoC}$ obtained from the edge server; and \textit{(iv)} possible penalty $q^{EtoC}$ paid from the edge server.

%f4
%\begin{figure*}[t!]
%\centering
%\subfigure[]{\includegraphics[width=.33\linewidth]{fig4a}}\hfill
%\subfigure[]{\includegraphics[width=.33\linewidth]{fig4b}}\hfill
%\subfigure[]{\includegraphics[width=.33\linewidth]{fig4c}}\\

%\subfigure[]{\includegraphics[width=.335\linewidth]{fig4d}}\hfill
%\subfigure[]{\includegraphics[width=.33\linewidth]{fig4e}}\hfill
%\subfigure[]{\includegraphics[width=.33\linewidth]{fig4f}}
%%\subfigure[]{\includegraphics[width=.245\linewidth]{fig4g}}\hfill
%%\subfigure[]{\includegraphics[width=.245\linewidth]{fig4h}}
%\caption{Performance evaluation from short-term perspective via simulating 100 trading.}
%\end{figure*}

%f5
\begin{figure*}[h!t]
\centering
\subfigure[]{\includegraphics[width=.33\linewidth]{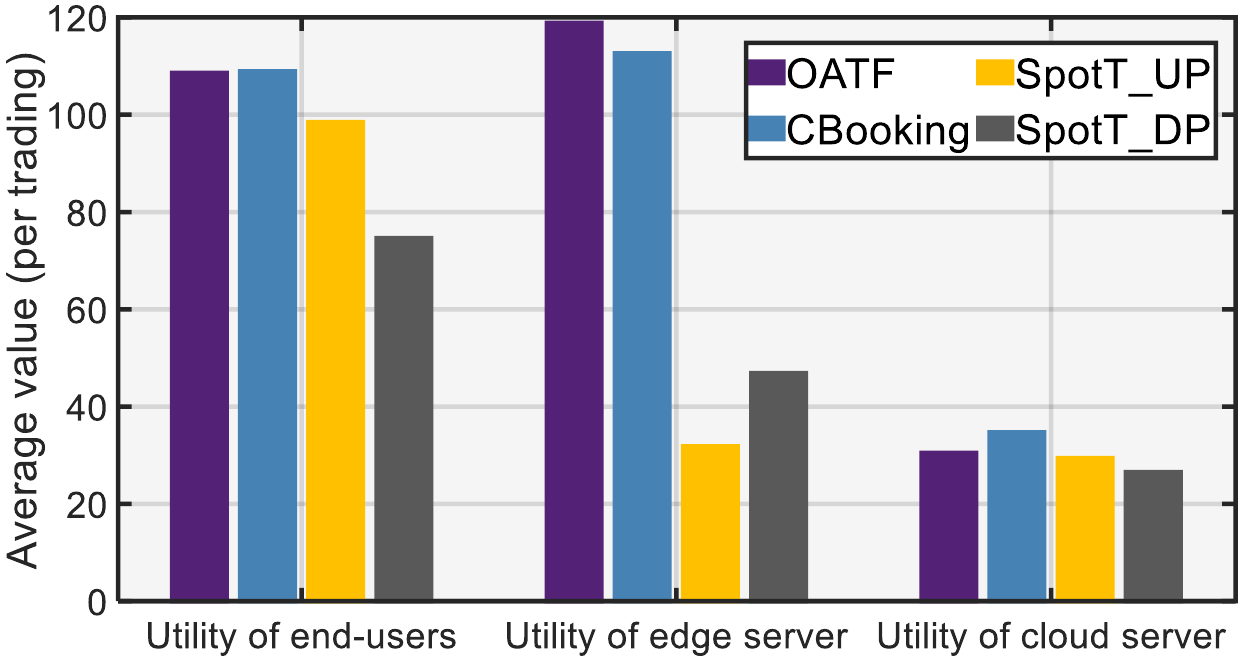}}\hfill
\subfigure[]{\includegraphics[width=.33\linewidth]{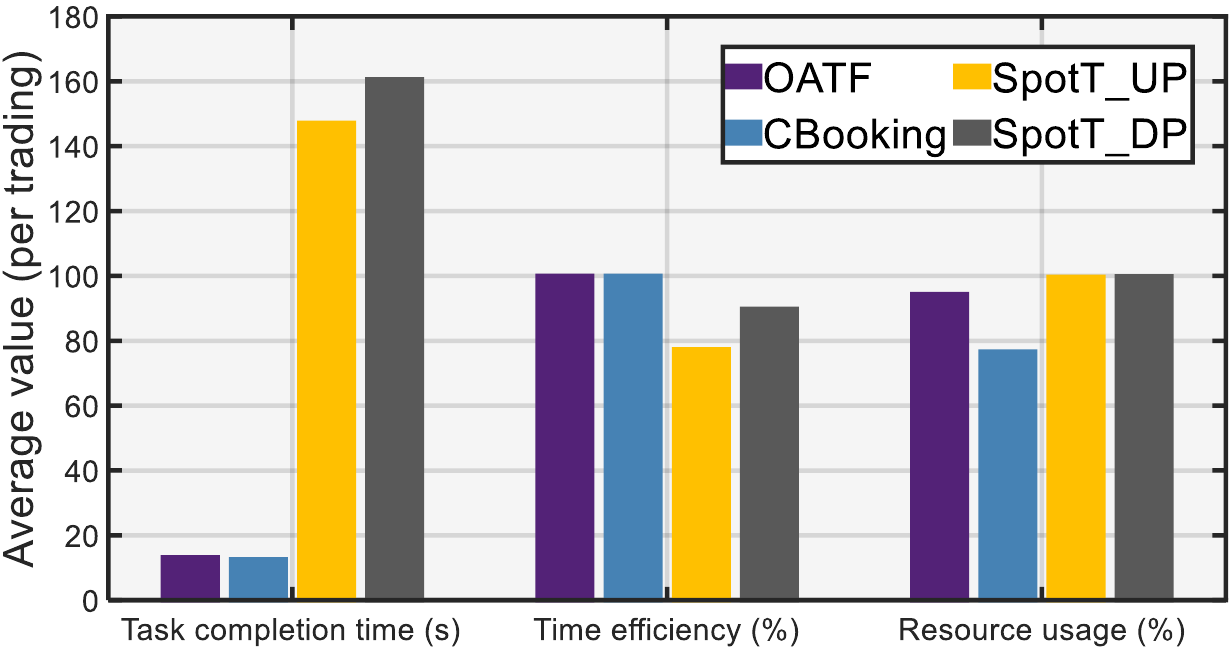}}\hfill
\subfigure[]{\includegraphics[width=.327\linewidth]{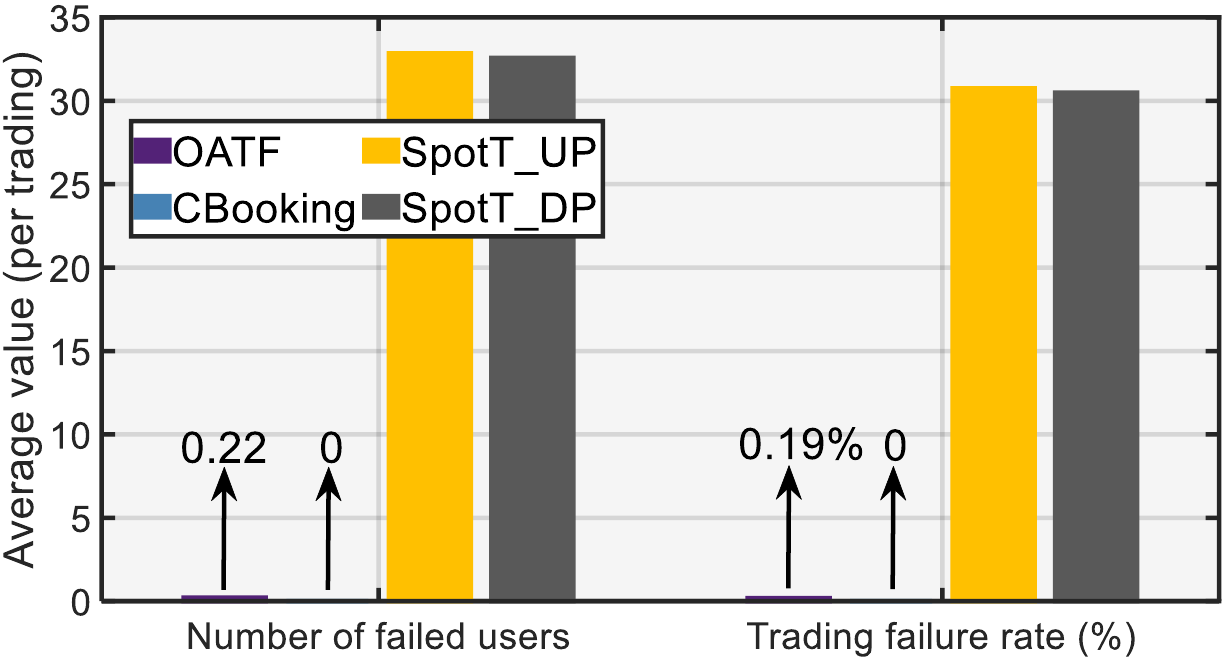}}
\caption{Performance evaluation from long-term perspective via simulating 5000 trading.}
\end{figure*}

\subsection{Analysis of Key Issues}
Analysis of key issues mentioned in the previous section based on the above models are discussed hereafter. First, before practical trading, the design of contracts, as well as overbooking rate is formulated as a multi-objective optimization (MOO) problem ($r^{User}=|\mathbb{U}|\times r^{user}$), aiming to maximize the expected utilities of end-users $\text{E}\left[\sum_{x=1}^{x=|\mathbb{U}|}\mathcal{U}^{u_x}\right]$, edge server $\text{E}\left[\mathcal{U}^{Edge}\right]$, and cloud server $\text{E}\left[\mathcal{U}^{Cloud}\right]$, while meeting tolerable risks as major constraints. 
Specifically, each end-user $u_x$ considers two key risks: \textit{(i)} the risk of receiving a negative utility, which is defined as the probability that $\mathcal{U}^{u_x}$ is less than or equal to 0; and \textit{(ii)} the risk of failing to acquire resources due to overbooking, which is calculated by the probability that conditions $\alpha_{x^\prime}=1$ and $r^{user}\sum_{x\neq x^\prime}\alpha_x>r^{Edge}+r^{Backup}-r^{User}$ are both met. Namely, the overall resources offered by the edge and cloud server fail to afford the demand of end-users, where some users thus have to process their applications locally. Similarly, the edge server is facing two major risks: \textit{(i)} the risk of obtaining an unsatisfying utility as defined by the probability that $\mathcal{U}^{Edge}$ is less than its expectation $\text{E}\left[\mathcal{U}^{Edge}\right]$; and \textit{(ii)} the risk of resource underutilization, which is reflected by the probability that resource usage stays below a certain rate (mainly caused by an improper overbooking rate). In addition, cloud server is undergoing the risk represented by the probability that the value of $\mathcal{U}^{Cloud}$ is smaller than or equal to $\text{E}\left[\mathcal{U}^{Cloud}\right]$. Apparently, all the above-mentioned risks should be well controlled within a certain range (e.g., each probability should not exceed a threshold. e.g., 30\%).

%(see MOO problem $\bm{\mathcal{F}}$ given by the following (1)). 
% \begin{align*}
%\bm{\mathcal{F}}:
% \begin{cases}
%\argmax\limits_{\mathbb{M}^{Edge}}\text{E}\left[\sum_{x=1}^{x=|\mathbb{U}|}\mathcal{U}^{u_x}\right]\\
%\argmax\limits_{\mathbb{M}^{Edge},\mathbb{M}^{Cloud}}\text{E}\left[\mathcal{U}^{Edge}\right]\\
%\argmax\limits_{\mathbb{M}^{Cloud}}\text{E}\left[\mathcal{U}^{Cloud}\right]
% \end{cases}\tag{1}
% \end{align*}
%~~~~~~~~~~~~~~~~~~~~~~~\text{s.t.} \textit{Tolerable risks} 

The proposed MOO problem faces difficulties to be solved directly by traditional algorithms such as the weighted sum method and $\epsilon$-constrained method, since it involves non-convex objective functions and complicated probabilistic constraints. 
To this end, a two-way multilateral negotiation scheme is designed that alternatively optimizes expected utilities, 
%$\text{E}\left[\sum_{x=1}^{x=|\mathbb{U}|}\mathcal{U}^{u_x}\right]$, $\text{E}\left[\mathcal{U}^{Edge}\right]$ and $\text{E}\left[\mathcal{U}^{Cloud}\right]$, 
while meeting acceptable risks. Specifically, "two-way" indicates that the edge server has to communicate with both end-users and the cloud server. Fig. 3 shows a diagram of how the proposed negotiation scheme is implemented among different parties, to reach the final agreement on forward contracts. As illustrated by Fig. 4, the edge server first starts a quotation process (step 1), while end-users can determine the acceptable range of $r^{user}$ (step 2) under tolerable risks under given the current price and default clause associated with contract $\mathbb{M}^{Edge}$. Then, the cloud server starts its quotation process, where the edge server decides the acceptable range of $r^{Backup}$, according to the price and default clause (step 4) related to both contracts 
%$\mathbb{M}^{Edge}$ and $\mathbb{M}^{Cloud}$
, under tolerable risks. Based on which, the cloud server determines feasible values of $r^{Backup}$ (step 5). When the three parties reach a consensus (step 6), end-users select the optimal pair of $r^{user}$ and $r^{Backup}$ that can maximize their expected utilities, which are regarded as candidate contract term, together with the corresponding prices and penalties (step 7). After the quotations of both edge and cloud servers are completed, the edge server chooses the final contract terms for both forward contracts $\mathbb{M}^{Edge}$ and $\mathbb{M}^{Cloud}$, to maximize its expected utility. Apparently, the edge server should understand both end-users and could server, which calls for two-way communication.

In particular, the corresponding computational complexity relies the total number of quotations of edge and cloud servers, which can be denoted by $O(Number~of~quotations)$. Notably, the overhead of OATF will only be incurred during contract negotiation (e.g. 10:00am in Fig. 1), while responsive and cost-effective computing services can be delivered directly during each practical trading, thanks to pre-signed contracts.

Then, each practical trading will proceed among participants on the basis of pre-determined forward contracts. Since end-users are i.i.d, FCFS mechanism is utilized for both compensated user selection and application transfer decision in this case study, which is also fair in real-world trading market.

\subsection{Results and Evaluation}

Simulation results associated with the proposed case study are carefully analyzed mainly from the long-term perspective, comparing with representative baseline methods listed below. 

\noindent
$\bullet$ \textbf{Conventional resource booking method (CBooking)} that doesn't allow the amount of resources reserved for end-users $r^{User}$ to exceed the available resource supply $r^{Edge}+r^{Backup}$.

\noindent
$\bullet$ \textbf{Spot trading method} where each practical trading is performed relying on the current market/network conditions (e.g., the current value of $\alpha_x$, $\gamma_x$ and $\beta$). Specifically, both uniform and differential pricing rules~\cite{4}~are considered, as abbreviated to ``SpotT\_UP'' and ``SpotT\_DP'', respectively.
%
%{\color{red}{Simulations are conducted via MATLAB R2020b platform on desktop computer with Intel Core i7–4770 3.40 GHz CPU and 16.0 GB RAM. }}

Major parameters are set as follows: $a=76\%$, $y_1=100$, $y_2=400$, $r^{Edge}=197$, $r^{Cloud}=600$, $\mathbb{U}=137$, $e^t=550\text{mW}$ ($e^t$ indicates the transmission power, while $e^t\gamma_x$ represents the received SNR of AP which is roughly within [17dB, 23dB]), $n=5$. Data size of each application is set to be 1Mb, while the corresponding required computing resources can be calculated as $600\times1024\text{bit}\times{1024}^2 \text{CPU cycles}$. Bandwidth of the channel between each end-user and the nearby AP (connected to edge server) is set by 6MHz. Besides, risks are controlled within interval $[20\%,40\%]$.
Since onsite decision-making relies heavily on the end-to-end (E2E) delay of wireless communication channels, a random variable $\tau_x$ (ms) is applied for each end-user in this simulation to describe E2E delay of the wireless channel between end-user $u_x$ and the nearby AP, which follows a uniform distribution denoted by $\tau_x\sim \text{U}\left(2,15\right)$~\cite{3}. 

%Fig. 4 illustrates short-term performance comparison upon having 100 practical trading. Specifically, although the proposed OATF mechanism sometimes obtain slightly lower resource usage than spot trading (since each spot trading is relying on the current resource supply/demand) as shown in Fig. 4(f), it generally achieves mutually beneficial utilities of the three parties, while facilitating faster application completion and better time efficiency than spot trading methods, especially comparing with Spot\_UP method as shown by Figs. 4(a)-4(e), thanks to the pre-determined forward contracts. CBooking method obtains similar performance on participants' utilities, application completion and time efficiency, and sometimes slightly better than OATF (e.g., Figs. 4(c)-4(d)), {\color{blue}{as benefited by more sufficient backup resources; however, is suffering from unsatisfying resource usage mainly caused by "no shows", as depicted in Fig. 4(f).}}

Evaluations upon considering the average value (per trading) of different indicators are shown in Fig. 5, via simulating 5000 trading. In Fig. 5(a), the proposed OATF mechanism outperforms Spot\_UP and Spot\_DP on the average value of participants' utilities, since onsite users have to spend extra time/energy on negotiating a trading agreement with the edge server which directly reduce the usable time for actual computing service delivery and thus time efficiency (also see Fig. 5(b)). Besides, onsite edge server may suffer from insufficient resource supply (e.g., cloud resources are occupied by other customers) and off-peak period (e.g., large number of users are absent from a trading).Besides, although overbooking can better handle dynamic resource demands, it also faces risks. This has led to similar performance of OATF in comparison with CBooking in Fig. 5(a), mainly caused by risks due to the random and unpredictable nature.  
Namely, the studied OATF relies on a risk and opportunity coexisting resource market, which is closer to real-world networks. For example, in Fig. 5(a), the cloud server’s utility of CBooking slightly outperforms that of OATF mainly because that overbooking may cause larger refunds paid to other customers of the cloud server (e.g., parameter $\beta$). 
Fortunately, the proposed OATF mechanism can get far better average resource usage rate than CBooking as illustrated by Fig. 5(b), since a feasible overbooking rate is encouraged that greatly supports dynamic resource demands. 

Fig. 5(c) depicts the average trading failures and the relevant rate, where all the trading in CBooking are successful owing to that the amount of promissory reserved resources for end-users in forward contracts equals to resource supply. In addition, the proposed OATF mechanism enables roughly 99.1\% and 99.3\% performance improvement on the number of failed users, and failure rate, rather than Spot\_UP and Spot\_DP, respectively, due to that onsite end-users may undergo insufficient resources and more severe competition, according to the current market/network conditions. For example, an end-user $u_x$ with large $\gamma_x$ can afford a higher resource price than $u_{x^\prime}$ with a small $\gamma_{x^\prime}$, which may lead to a failure although $u_{x^\prime}$ has spent time on negotiating with the edge server.

%6
\section{Conclusion and Future Direction}
\noindent
Big data generated by countless IoT devices highly requests real-time and cost-effective application processing, which calls for additional requirements on resource provisioning techniques in dynamic networks. This article investigates a novel overbooking-enabled forward trading mechanism called OATF, under device-edge-cloud network architecture with various uncertainties. OATF relies on pre-determined forward contracts negotiated among end-users and the edge server, as well as between the edge server and cloud server in advance, which will be fulfilled accordingly during each future practical trading. Specifically, a certain overbooking rate is encouraged that greatly supports the substantial resource utilization under dynamic resource demand. Framework and key issues associated with the proposed OATF mechanism are carefully analyzed, based on which, an interesting case study is investigated via specific mathematical models. Comprehensive simulation results illustrate that the proposed OATF mechanism can achieve commendable performance on various evaluation indicators such as time efficiency and resource usage, in comparison with conventional trading methods.

%Since OATF offers a commendable reference for the future development of resource trading in mobile networks, s

Several interesting research directions are worthy of consideration. For example, smart and alterable rights/obligations rather than fixed contract terms are noteworthy to study via adopting machine learning approaches, to better capture the dynamics and unpredictable nature of resource trading market. Moreover, multiple edge servers can be considered, where competition and cooperation among them stand for another future direction.

\section*{Acknowledgement}

\noindent
This work was supported in part by the Fundamental Research Funds for the Central Universities under No. 20720220104, the Discovery Program of Natural Sciences and Engineering Research Council of Canada (NSERC) under Grant RGPIN2018-06254, and the Canada Research Chair Program.

\section*{Biography}
\begin{IEEEbiographynophoto}{Minghui Liwang} (minghuilw@xmu.edu.cn) received her Ph. D. degree in the School of Informatics, Xiamen University, China, in 2019. She is now an assistant professor in the School of Informatics, Xiamen University, China. Her research interests include wireless communication systems, cloud/edge/service computing, economic models and applications.
\end{IEEEbiographynophoto}

\begin{IEEEbiographynophoto}{Xianbin Wang}(xianbin.wang@uwo.ca) is a
professor and Tier 1 Canada Research Chair at Western University. He is a Fellow of Canadian Academy of Engineering, a Fellow of IEEE, a Fellow of the Engineering Institute of Canada, and an IEEE Distinguished Lecturer. His current research interests include 5G and beyond, intelligent wireless communications,
Internet of Things, and communications security. 
\end{IEEEbiographynophoto}

\vfill

\end{document}